\documentclass{article}
\usepackage{frascatiphys}
\usepackage{graphicx}
\usepackage{xspace}
\usepackage{floatrow}
\usepackage{amsmath}
\usepackage{amsfonts}
\usepackage{amssymb}
\usepackage{calrsfs}
\usepackage[colorlinks=true,allcolors=blue]{hyperref}
\newfloatcommand{capbtabbox}{table}[][\FBwidth]
\newcommand{\epem}{e^+e^-}
\newcommand{\alphas}{\alpha_{\rm s}}
\newcommand{\sqrts}{\sqrt{\rm s}}
\newcommand{\qqbar}{q\overline{q}}
\newcommand{\bbbar}{b\overline{b}}
\newcommand{\ttbar}{t\overline{t}}
\newcommand{\mZ}{m_{_\mathrm{\rm Z}}}
\newcommand{\mW}{m_{_\mathrm{\rm W}}}

\newcommand*{\eg}{e.g.\@\xspace}
\newcommand*{\ie}{i.e.\@\xspace}
\newcommand*{\cm}{c.m.\@\xspace}
\def\cO#1{{{\cal{O}}}\left(#1\right)}
\def\ttt#1{\texttt{\tiny #1}}

\begin{document}
\title{Physics case of FCC-ee} 
\author{
David d'Enterria       \\
{\em CERN, PH-EP Department, 1211 Geneva, Switzerland}
}
\maketitle
\baselineskip=11.6pt
\begin{abstract}
The physics case for electron-positron beams at the Future Circular Collider (FCC-ee) is succinctly
summarized. The FCC-ee core program involves $\epem$ collisions at 
$\sqrts$~=~90, 160, 240, and 350~GeV with multi-ab$^{-1}$ integrated luminosities, 
yielding about 10$^{12}$ Z bosons, 10$^{8}$ W$^+$W$^-$ pairs, 10$^{6}$ Higgs bosons and 4$\cdot$10$^{5}$
$\ttbar$ pairs per year. The huge luminosities combined with $\cO{100~\rm keV}$ knowledge of the \cm\ energy
will allow for Standard Model studies at unrivaled precision. Indirect constraints on new physics 
can thereby be placed up to scales $\Lambda_{_{\rm NP}} \approx$~7 and 100~TeV for particles coupling
respectively to the Higgs and electroweak bosons. 
\end{abstract}
\baselineskip=14pt
%

\section{Introduction}
\begin{figure}[htb]
\centering
\includegraphics[scale=0.63]{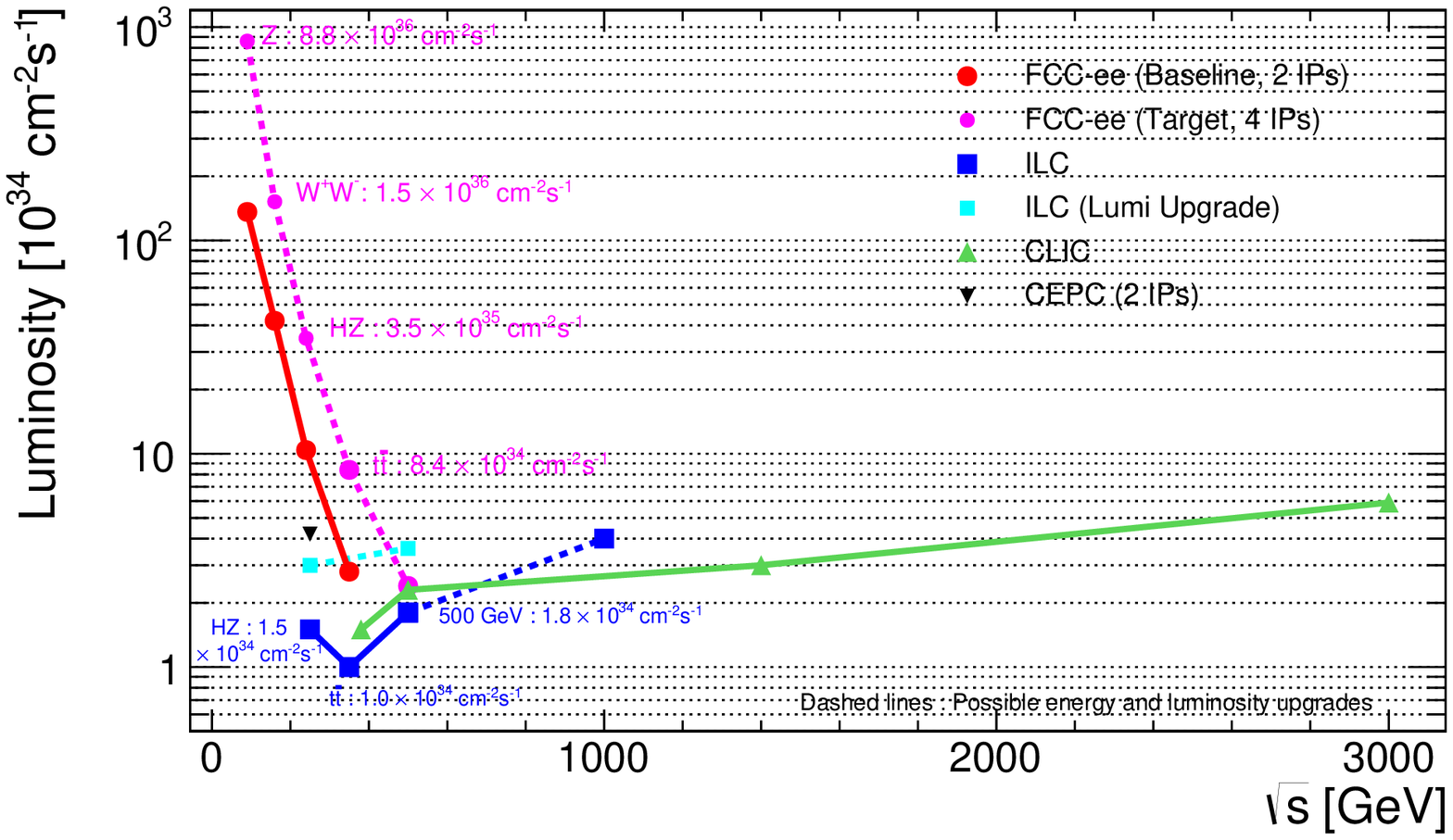}
\caption{\it Target luminosities as a function of center-of-mass energy for future circular (FCC-ee, CEPC)
  and linear (ILC, CLIC) $\epem$  colliders under consideration.}
\label{fig:ee_lumi}
\end{figure}

The Standard Model (SM) of particle physics is a renormalizable quantum field theory encoding our knowledge of
the fundamental particles and their (electroweak and strong) interactions. Despite its tremendous success to
describe many phenomena with high accuracy for over 40 years --including the recent experimental confirmation
of the existence of its last missing piece, the Higgs boson-- fundamental questions remain open (such as the
small Higgs boson mass compared to the Planck scale, dark matter, matter-antimatter asymmetry, neutrino masses,...)
which may likely {\it not} be fully answered through the study of proton-proton collisions at the Large Hadron
Collider (LHC). Notwithstanding their lower center-of-mass energies, high-energy $\epem$ colliders feature several
advantages in new physics studies compared to hadronic machines, such as direct model-independent searches of
new particles coupling to Z*/$\gamma$* with masses up to $m\approx\sqrts/2$, and clean experimental 
environment with  initial and final states very precisely known theoretically, (\ie\ well understood
backgrounds without ``blind spots'' of p-p searches).
Combined with high-luminosities, an $\epem$ collider 
can thus provide access to studies with $\delta X$ precision at the permille level, allowing indirect
constraints to be set on new physics up to very-high energy scales 
$\Lambda_{_{\rm NP}}\propto$~(1~TeV)/$\sqrt{\delta X}$. 
Plans exist to build future circular (FCC-ee\cite{FCCee}, CEPC\cite{CEPC}) and/or linear (ILC\cite{ILC},
CLIC\cite{CLIC}) $\epem$ colliders (Fig.~\ref{fig:ee_lumi}).
The advantages of circular machines are (i) their much higher luminosity below $\sqrts\approx$~400~GeV
(thanks to much larger collision rates, adding continuous top-up injection to compensate for luminosity 
burnoff), (ii) the possibility to have several interaction points (IPs), and (iii) a precise measurement of
the beam energy E$_{\rm beam}$ through resonant transverse depolarization\cite{Koratzinos:2015hya}. Linear 
colliders, on the other hand, feature (i) much larger $\sqrts$ reach (circular colliders are not competitive
above $\sqrts\approx$~400~GeV due to synchroton radiation scaling as E$_{\rm beam}^4$/R), and (ii) easier
longitudinal beam polarization. At the FCC (with a radius R~=~80--100~km), $\epem$ collisions present clear
advantages with respect to LEP (\eg\ $\times$10$^4$ more bunches, and $\delta E_{\rm beam}\approx$~$\pm$0.1~MeV 
compared to $\pm$2~MeV) and to ILC (crab-waist optics scheme, up to 4 IPs) yielding luminosities
$\times$(10$^4$--10) larger in the $\sqrts$~=~90--350~GeV range\cite{Zimmermann}.
Table~\ref{tab:runs} lists the target FCC-ee luminosities, and the total number of events collected at each $\sqrts$ 
obtained for the following cross sections (including initial state radiation, and smearings due to beam-energy spreads): 
$\sigma_{\rm \epem\to Z}=$~43~nb, $\sigma_{\rm \epem\to H}=$~0.29~fb, $\sigma_{\rm \epem\to W^+W^-}=$~4~pb, 
$\sigma_{\rm \epem\to HZ}=$~200~fb, $\sigma_{\rm \epem\to \ttbar}=$~0.5~pb, and $\sigma_{\rm \epem\to VV \to
  H}=$~30~fb. With these target luminosities, the completion of the FCC-ee core physics program (described in
the next sections) requires 10 years of running.

\renewcommand\arraystretch{1.3}
\begin{table}[htpb!]
\centering
\scriptsize
 \begin{tabular}{|l|c|c|c|c|c|c|} \hline 
 $\sqrts$ (GeV): & 90 (Z) & 125 (eeH) & 160 (WW) & 240 (HZ) & 350 ($\ttbar$) &  350 ($\rm VV\to H$) \\ \hline\hline
 $\cal{L}/$IP (cm$^{-2}$\,s$^{-1}$) & 2.2$\cdot$10$^{36}$ & 1.1$\cdot$10$^{36}$ & 3.8$\cdot$10$^{35}$ & 8.7$\cdot$10$^{34}$ & 2.1$\cdot$10$^{34}$ & 2.1$\cdot$10$^{34}$ \\ 
 $\cal{L}_{\rm int}$ (ab$^{-1}$/yr/IP) & 22 &  11 & 3.8 & 0.87 & 0.21 & 0.21 \\ 
 Events/year (4 IPs) & 3.7$\cdot$10$^{12}$ & 1.3$\cdot$10$^{4}$ & 6.1$\cdot$10$^{7}$ & 7.0$\cdot$10$^{5}$ & 4.2$\cdot$10$^{5}$ & 2.5$\cdot$10$^{4}$ \\ 
 Years needed (4 IPs) &  2.5 &  1.5 & 1 & 3 & 0.5 & 3 \\ \hline
 \end{tabular}
 \caption{\it Target luminosities, events/year, and years needed to complete the W, Z, H and
   top programs at FCC-ee. [{\scriptsize $\cal{L}$~=~10$^{35}$~cm$^{-2}$\,s$^{-1}$ corresponds to
$\cal{L}_{\rm int}$~=~1~ab$^{-1}$/yr for 1 yr = 10$^7$ s}].}
 \label{tab:runs}
\end{table}

\section{Indirect  constraints on BSM via high-precision Z, W,  top physics}

Among the main goals of the FCC-ee is to collect multi-ab$^{-1}$ at $\sqrts\approx$~91~GeV (Z pole), 160~GeV
(WW threshold), and 350~GeV ($\ttbar$ threshold) in order to measure  with unprecedented precision key
properties of the W and Z bosons and top-quark, as well as other fundamental parameters of the SM.
The combination of huge data samples available at each $\sqrts$, and the precise knowledge of the \cm\ energy 
leading to very accurate threshold scans, allows for improvements in their experimental precision by factors
around $\times$25 (dominated by systematics uncertainties) compared to the current state-of-the-art
(Table~\ref{tab:SM})\cite{Tenchini:2014lma}. In many cases, the dominant uncertainty will be of theoretical
origin, and developments in the calculations are needed in order to match the expected experimental uncertainty. 
The FCC-ee experimental precision targets are \eg\ $\pm$100~keV for $\mZ$, $\pm$500~keV for $\mW$,
$\pm$10~MeV for $m_\mathrm{t}$, a relative statistical uncertainty of the order of 3$\cdot$10$^{-5}$ for the
QED $\alpha$ coupling (through muon forward-backward asymmetries above and below the Z
peak)\cite{Janot:2015gjr}, 1-permille for the QCD coupling $\alphas$ (through hadronic Z and W
decays)\cite{d'Enterria:2015toz}, and $10^{-3}$ on the electroweak top couplings
$F^{\gamma\,t,Z\,t}_\mathrm{1V,2V,1A}$ (through differential distributions in $\epem\to\ttbar\to 
\ell\nu\qqbar\bbbar$)\cite{Janot:2015yza}.

\begin{figure}[htbp!]
\centering
\includegraphics[width=0.47\columnwidth,height=4.5cm]{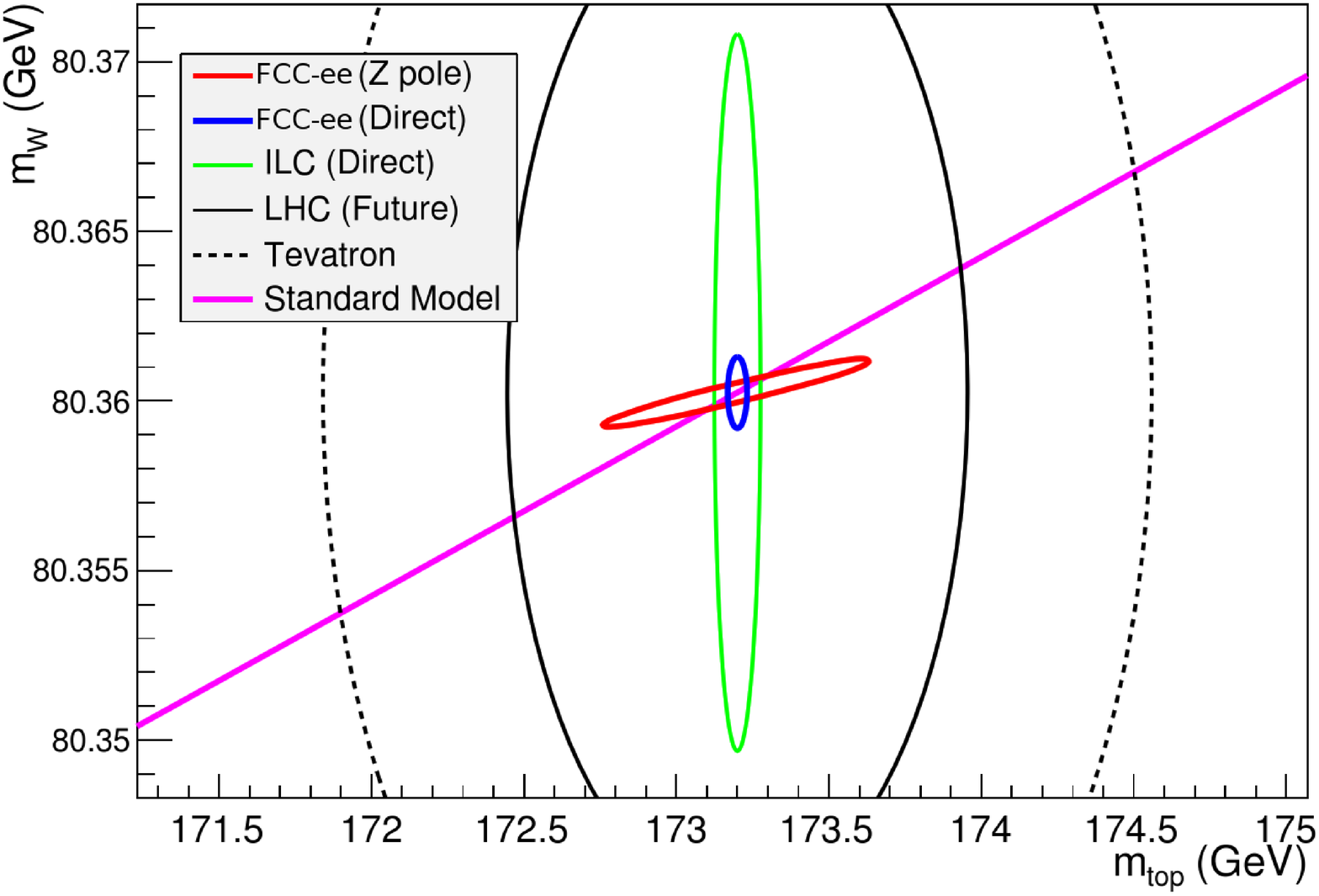}\hspace{0.2cm}
\includegraphics[width=0.47\columnwidth,height=4.5cm]{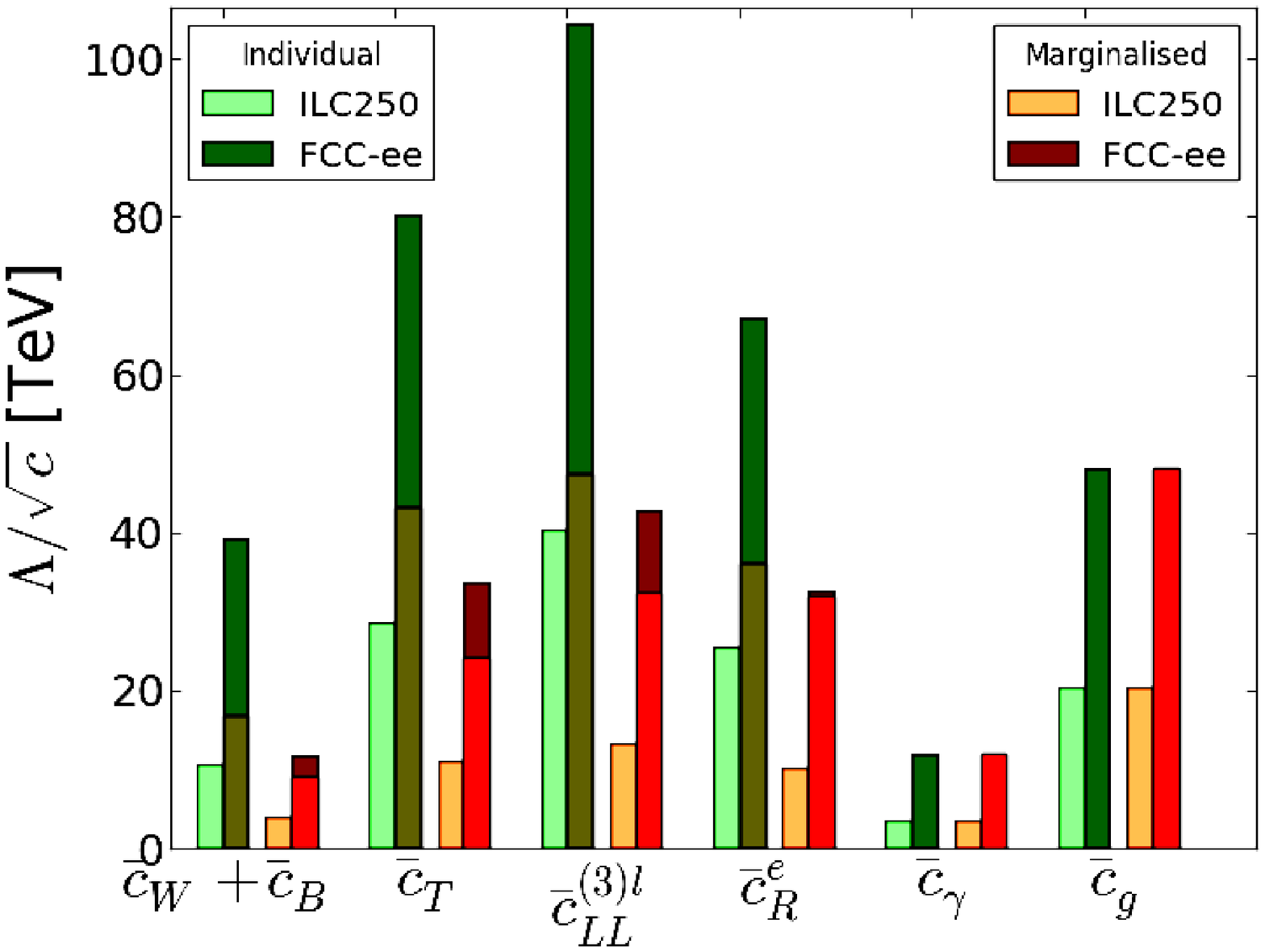}
\caption{\it Left: 68\% C.L. limits in the $m_\mathrm{t}$--$\mW$ plane at FCC-ee and other colliders\cite{FCCee}.
 Right: Energy reaches for dim-6 operators sensitive to EWPT, obtained from precision measurements at FCC-ee
 and ILC\cite{Ellis:2015sca}.}
\label{fig:FCCee_SM}
\end{figure}

\renewcommand\arraystretch{1.3}
\begin{table}[htpb!]
\centering
\scriptsize
 \begin{tabular}{|@{}c@{}|@{}c@{}|@{}c@{}|@{}c@{}|@{}c@{}|@{}c@{}|} \hline 
 Observable & Measurement & Current precision & \;FCC-ee stat.\; & \;Possible syst.\; & Challenge \\ \hline\hline
 $\mZ$ (MeV) & Z lineshape & $91187.5 \pm 2.1$ & 0.005 & $<0.1$ & QED corrs. \\ 
 $\Gamma_{_\mathrm{Z}}$ (MeV) & Z lineshape & $2495.2 \pm 2.3$ & 0.008 & $<0.1$ & QED corrs. \\ 
 $R_\ell$ & Z peak & $20.767 \pm 0.025$ & 0.0001 & $<0.001$ & QED corrs. \\ 
 $R_\mathrm{b}$ & Z peak & $0.21629 \pm 0.00066$ & 0.000003 & $<0.00006$ & $g\rightarrow \mathrm{b\bar{b}}$ \\ 
 $A_{_\mathrm{FB}}^{\mu\mu}$ & Z peak & $0.0171 \pm 0.0010$ & 0.000004 & $<0.00001$ & $E_\mathrm{beam}$ meas. \\ 
 $N_\nu$ & Z peak & $2.984 \pm 0.008$ & 0.00004 & $0.004$ & Lumi meas. \\ 
 $N_\nu$ & $\mathrm{e^+e^- \rightarrow \gamma\, Z(inv.)}$ & $2.92 \pm 0.05$ & 0.0008 & $<0.001$ & -- \\ 
 $\alphas(\mZ)$ & $R_\ell,\sigma_\mathrm{had},\Gamma_{_\mathrm{Z}}$ & $0.1196 \pm 0.0030$ & 0.00001 &
 $0.00015$ & New physics \\ 
 $\;1/\alpha_{_\mathrm{QED}}(\mZ)\;$ & \;$A_{_\mathrm{FB}}^{\mu\mu}$ around Z peak\; & $128.952 \pm 0.014$ & 0.004 & 0.002 & EW corr. \\ \hline
 $\mW$ (MeV) & WW threshold\ scan & $80385 \pm 15$ & 0.3 & $<1$ & QED corr. \\ 
 $\alphas(\mW)$ & $B^{^{\rm W}}_\mathrm{had}$ & $B^{^{\rm W}}_\mathrm{had} = 67.41 \pm 0.27$ & 0.00018 & $0.00015$ & \;CKM matrix\; \\ \hline
 $m_\mathrm{t}$ (MeV)& threshold scan & $173200 \pm 900$ & 10 & 10& QCD \\ 
 $F^{\gamma\,t,Z\,t}_\mathrm{1V,2V,1A}$& d$\sigma^{\ttbar}$/dx\,d$\cos(\theta)$ & \;4\%--20\% (LHC-14 TeV)\; &
 (0.1--2.2)\% & (0.01--100)\% & -- \\ \hline
 \end{tabular}
 \caption{\it Examples of achievable precisions in representative Z, W and top measurements.} 
 \label{tab:SM}
\end{table}
Figure~\ref{fig:FCCee_SM} shows limits in 
the W-mass vs. top-mass plane (left), and the energy reaches of a subset of dimension-6 operators of an Effective
Field Theory of the SM parametrizing possible new physics (right)\cite{Ellis:2015sca}.
Such measurements impose unrivaled constraints on new weakly-coupled physics. Whereas electroweak precision
tests (EPWT) at LEP bound any BSM physics to be above $\Lambda_{_{\rm NP}}\gtrsim$~7~TeV, FCC-ee would reach 
up to $\Lambda_{_{\rm NP}}\approx$~100~TeV.

\section{Indirect  constraints on BSM via high-precision Higgs physics}

\begin{figure}[htpb!]
\centering
\begin{floatrow}
\ffigbox{%
}{%
\includegraphics[height=7.2cm,width=6.cm]{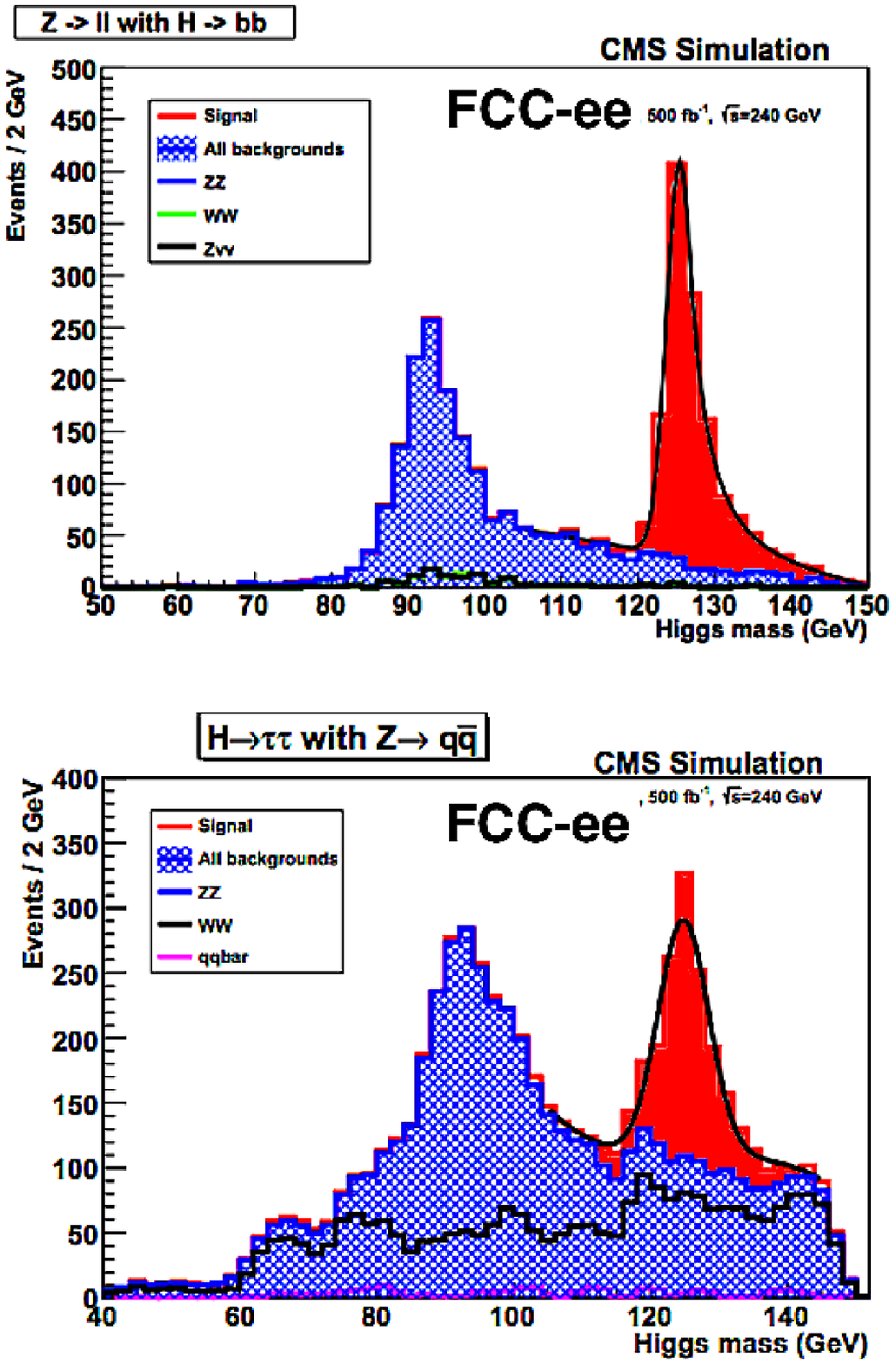}
  \caption{Distribution of recoil mass against $\rm Z\to\ell\ell$ (top) and  $\rm Z\to\qqbar$ (bottom) in the 
$\epem\to\rm HZ$ process with H$\to\bbbar$ (top) and  H$\to\tau\tau$ (bottom).
\label{fig:HZ}}%
}
\capbtabbox{%
\begin{tabular}{|c|c|c|}\hline
\hline Observable & 240 GeV & 240+350 GeV \\ \hline
\hline $g_{\rm HZZ}$ & 0.16\% & 0.15\% \\ 
\hline $g_{\rm HWW}$ & 0.85\% & 0.19\% \\ 
\hline $g_{\rm Hbb}$ & 0.88\% & 0.42\% \\ 
\hline $g_{\rm Hcc}$ & 1.0\% & 0.71\% \\ 
\hline $g_{\rm Hgg}$ & 1.1\% & 0.80\% \\ 
\hline $g_{{\rm H}\tau\tau}$ & 0.94\% & 0.54\% \\ 
\hline $g_{{\rm H}\mu\mu}$ & 6.4\% & 6.2\% \\ 
\hline $g_{{\rm H}\gamma\gamma}$ &1.7\% & 1.5\% \\ \hline
\hline $\Gamma_{\rm tot}$ & 2.4\% & 1.2\% \\ \hline
\hline ${\rm BR}_{\rm inv}$ & 0.25\% & 0.2\% \\ \hline
\hline ${\rm BR}_{\rm exo}$ & 0.48\% & 0.45\% \\ \hline
\end{tabular} 
}{%
  \caption{Expected model-independent uncertainties on Higgs couplings, total width,
  and branching ratios into invisible and exotic particles (invisible or not)\cite{FCCee}.\label{tab:Higgs}}%
}
\end{floatrow}
\end{figure}

In the range of \cm\ energies covered by the FCC-ee, Higgs production peaks at $\sqrts\approx$~240~GeV 
dominated by Higgsstrahlung ($\epem\to{\rm HZ}$), with some sensitivity  also to
vector-boson-fusion ($VV\to {\rm H}\;\epem,\nu\nu$) and the top Yukawa coupling ($\epem\to\ttbar$ with a virtual
Higgs exchanged among the top quarks) at $\sqrts$~=~350~GeV. The target total number of Higgs produced at the FCC-ee (4 IPs
combined, all years) amounts to 2.1~million at 240~GeV, 75\,000 in $VV\to{\rm H}$ at 350~GeV, and 19\,000
in s-channel $\epem\to{\rm H}$ at $\sqrts$~=~125~GeV (Table~\ref{tab:runs}).
With such large data samples, unique Higgs physics topics are accessible to study:
\begin{itemize}
\item High-precision model-independent determination of the Higgs couplings, total width, and exotic 
  and invisible decays (Table~\ref{tab:Higgs})\cite{FCCee}.
\item Higgs self-coupling through loop corrections in HZ production\cite{McCullough:2013rea}.
\item First-generation fermion couplings: (u,d,s) through exclusive decays H$\to V\gamma$
  ($V=\rho,\omega,\phi$)\cite{Kagan:2014ila}, and electron Yukawa through resonant $\epem\to\rm H$ at 
$\sqrts=m_{_{\rm H}}$\cite{DdE}. 
\end{itemize}
The recoil mass method in $\epem\to\rm HZ$ is unique to lepton colliders and allows for a
high-precision tagging of Higgs events irrespective of their decay mode (Fig.~\ref{fig:HZ}).
It provides, in particular, a high-precision ($\pm$0.05\%) measurement of $\sigma_{\rm \epem\to HZ}$ and,
therefore, of $g_{\rm HZ}^2$. From the measured value of $\sigma_{\rm \epem\to H(XX)Z}\propto\Gamma_{\rm H\to XX}$ and 
different known decays fractions $\Gamma_{\rm H \to XX}$, one can then obtain the total Higgs boson width
with $\cO{1\%}$ uncertainty combining different final states. The $\ell^+\ell^-{\rm H}$ final state and the
distribution of the mass recoiling against the lepton pair can also be used to directly measure the invisible
decay width of the Higgs boson in events where its decay products escape undetected. The Higgs boson invisible
branching fraction can be measured with an absolute precision of 0.2\%. If not observed, a 95\% C.L. upper
limit of 0.5\% can be set on this branching ratio\cite{FCCee}.    
In addition, loop corrections to the Higgsstrahlung cross sections at different center-of-mass energies are
sensitive to the Higgs self-coupling\cite{McCullough:2013rea}. The effect is tiny but visible at FCC-ee thanks
to the extreme precision achievable on the $g_{\rm HZ}$ coupling. Indirect and model-dependent limits on the
trilinear $g_{\rm HHH}$ can be set with a $\cO{70\%}$ uncertainty, comparable to that expected at HL-LHC.\\


The large Higgs data samples available also open up to study exotic (\eg\ flavour-violating Higgs)
and very rare SM decays. First- and second-generation couplings to fermions are accessible via exclusive
decays $\rm H\to V\gamma$, for $V=\rho,\omega,\phi$, with sensitivity to the u,d,s quark
Yukawas\cite{Kagan:2014ila}. The $\rm H\to\rho\gamma$ channel appears the most promising with $\cO{50}$ events
expected. The low mass of the electron translates into a tiny $\rm H\to\epem$ branching ratio ${\rm BR}_{\rm
  \epem} = 5\cdot 10^{-9}$ which precludes any experimental observation of this decay mode and, thereby a
determination of the electron Yukawa coupling. The resonant s-channel production, despite its small cross
section\cite{Jadach:2015cwa}, is not completely hopeless and preliminary studies indicate that one could
observe it at the 5$\sigma$-level accumulating 75~ab$^{-1}$ at FCC-ee running at $\sqrts$~=~125~GeV with a
\cm\ energy spread commensurate with the Higgs boson width itself ($\approx$4~MeV) which requires beam
monochromatization\cite{DdE}.\\ 

Summarizing in terms of new physics constraints, deviations $\delta g_{\rm HXX}$ of the Higgs boson couplings to gauge
bosons and fermions with respect to the SM predictions can be translated into BSM scale limits through:
$\rm \Lambda_{_{\rm NP}}\gtrsim (1\,TeV)/\sqrt{(\delta g_{\rm HXX}/g_{\rm HXX}^{\rm SM})/5\%}$.
The expected 0.15\% precision for the most precise coupling, $g_{\rm HZZ}$, would thus set competitive bounds,
$\Lambda_{_{\rm NP}}\gtrsim$~7~TeV, on any new physics coupled to the scalar sector of the SM. 

\section{Direct constraints on BSM physics: dark matter and right-handed neutrinos}

\begin{figure}[htbp!]
\centering
\includegraphics[width=0.48\columnwidth,height=5.cm]{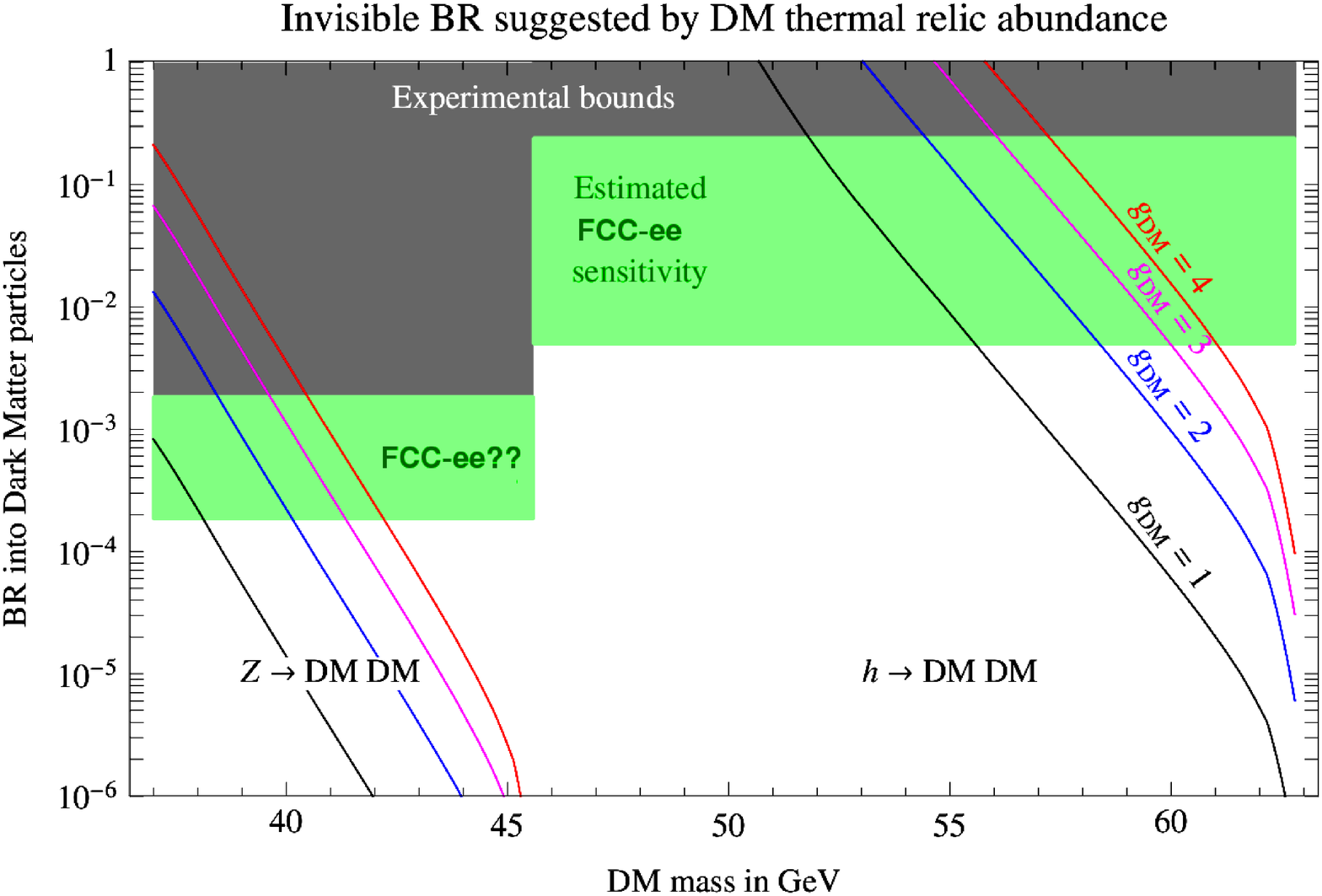}\hspace{0.1cm}
\includegraphics[width=0.49\columnwidth,height=5.cm]{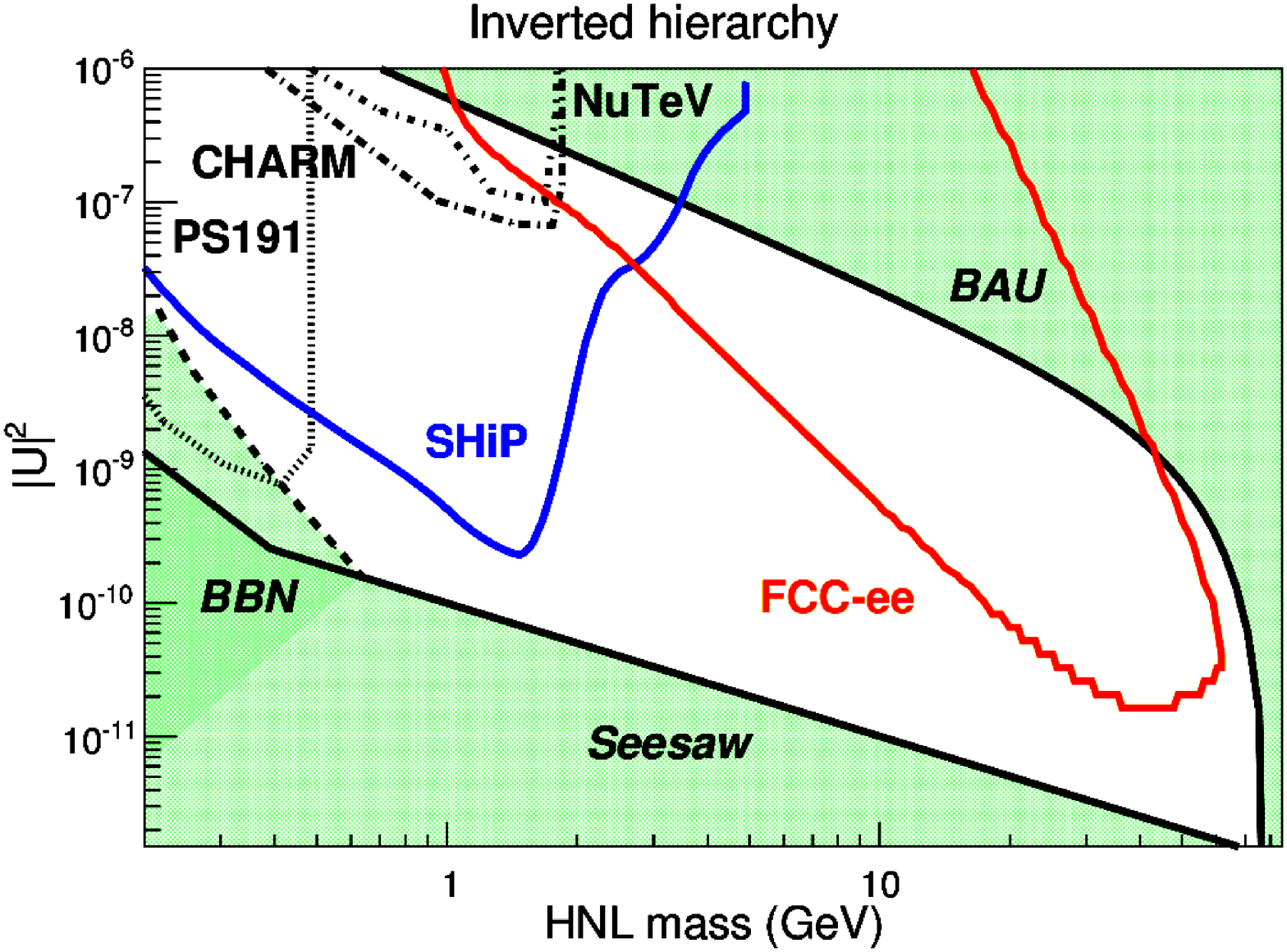}
\caption{\it Regions of sensitivity of FCC-ee for: (i) Z and H decays into DM in the 
$\rm BR_{_{Z,H\to\,DM\,DM}}$ vs. $\rm m_{_{DM}}$ plane (left)\cite{Strumia15}, and (ii) sterile neutrinos as a 
function of their mass and mixing to light neutrinos (inverted hierarchy) for 10$^{12}$ Z decays (right)\cite{Blondel:2014bra}.
}
\label{fig:FCCee_BSM}
\end{figure}

The precision electroweak and Higgs boson studies, summarized previously, not only impose
competitive constraints on new physics at multi-TeV scales but can also be interpreted in terms of
limits in benchmark SUSY models (CMSSM and NUHM1)\cite{FCCee}. Other studies exist that have
analyzed the impact of FCC-ee on other key BSM 
extensions such as \eg\ direct searches of dark matter (DM)\cite{Strumia15} and right-handed 
neutrinos\cite{Blondel:2014bra} through Z and H bosons rare decays. Figure~\ref{fig:FCCee_BSM} (left)
shows the limits in the plane (branching ratio, DM mass) for the decays $\rm Z,H\to DM\,DM$.
Measurements of the invisible Z and H widths are the best collider options to test DM lighter
than $m_{_{\rm Z,H}}/2$ that couples via SM mediators. Figure~\ref{fig:FCCee_BSM} (right) shows the
unrivaled limits that can be set in sterile neutrinos searches via decays $\rm Z\to N\nu_i$ with $\rm N\to
W^*\ell,Z^*\nu_j$ as a function of their mass and mixing to light neutrinos\cite{Blondel:2014bra}.

\vspace{-0.5cm}
\section{Summary}

Electron-positron collisions at $\sqrts \approx$~90--350~GeV at the FCC-ee provide unique means to address
many of the fundamental open problems in particle physics via high-precision studies of the W, Z, Higgs, and 
top-quark with permil-level uncertainties, thanks to the huge luminosities $\cO{1\rm{-}100}$~ab$^{-1}$ 
and the exquisite beam energy calibration. 
Such measurements set indirect constraints on BSM physics up to scales 
$\Lambda_{_{\rm NP}} \approx 7,100$~TeV for new particles coupling to the scalar and electroweak SM sectors,
respectively. Rare Higgs and Z bosons decays are sensitive to dark matter and
sterile neutrinos with masses up to $m_{_{\rm DM,HNL}}\approx$~60~GeV. 


%
\end{document}